\documentstyle[12pt]{article}
\advance\textheight by \topskip
\def\fun#1#2{\lower3.6pt\vbox{\baselineskip0pt\lineskip.9pt
  \ialign{$\mathsurround=0pt#1\hfil##\hfil$\crcr#2\crcr\sim\crcr}}}
\relax
\begin{document}
\begin{flushright}
\end{flushright}
\vskip 2 cm
\begin{center}
{\LARGE \bf The Deflation of $SU(3)_c$ 
at High Temperatures }\vskip 1.7cm

{\bf Afsar Abbas$^{\#}$, Lina Paria$^{\#}$ and
Samar Abbas$^{*}$ } \vskip 3 mm 
$\#$.Institute of Physics, Bhubaneswar - 751005,
Orissa, India \\
$*$. Department of Physics, Utkal University,
Bhubaneswar-751004, 
Orissa, India \footnote{ 
e-mail :afsar/lina/abbas@iopb.res.in }

\end{center}
\vskip 2 cm

{\centerline{\large\bf Abstract}}

\begin{quotation}
The ideas of ``local" and ``global" colour-singletness are not
well understood within QCD.
We use a group theoretical technique to project out the partition
function for a system of quarks, antiquarks and gluons to a particular
representation of the internal symmetry group $SU(3)_c$:
colour-singlet, colour-octet and colour 27-plet at finite temperature. 
For high temperatures and large size it is shown that 
colour-singlet is degenerate with colour-octet, colour 27-plet states etc. 
For the composite system 
it is shown that $ SU(3)_c $ appears to be a good symmetry only at 
low temperatures and at higher temperatures it gets submerged into a 
larger group $ U(12)_q \otimes U(12)_{\bar{q}} $ (2-flavour). At high 
enough temperatures this conclusion is model independent. This means that a
phase transition from the hadronic matter to the quark-gluon phase implies
a transition from the group $SU(3)_c$ to $U(12)_q \otimes U(12)_{\bar q}$.
Ideas of extensions beyond the standard model would have to be reviewed 
in the light of this result.

\end{quotation}

\newpage

\vskip 5 mm 

{\bf 1. INTRODUCTION } \\
\vskip 0.3 cm

Colour confinement is an experimentally well
established property of QCD at temperature T = 0. Though it has not been 
conclusively demonstrated in QCD it is universally believed to be true. 
Confinement is definitely not shown to be sine quo non of the non-abelian
QCD. Several model calculations indicate that indeed the 3-q and 
$ q\bar{q}$ colour-singlet states are more bound than for example the 
colour octet, decuplet, etc. representations \cite{mars}.
However one cannot simply throw away the higher colour representations
like octet as they may manifest themselves in specific situations like in
the multiquark systems \cite{afsar} etc. Recently the colour-octet 
contribution has also been shown to be significant during quarkonium production
in hadronic collisions \cite{bra}. The question we ask in this paper is 
what is the role of higher representations like 8-plet, 27-plet etc.
for the bulk QGP at high temperatures and their possible role in the
early universe QCD phase transition and also the QGP scenarios in the 
heavy ion collision setup. Below we shall demonstrate their
significance and the new insights that they bring in.

It is believed that in QCD ``transition from hadronic matter to
the quark-gluon matter is a transition from local colour confinement 
(on the scale of 1 fm) to global colour confinement" \cite{mull}. 
It is not understood as to what maintains long-range correlations
implicit in global colour-confinement, for example for sizes as large as
the quark stars \cite{glender}.
To better understand the role of the colour degree
of freedom we use the colour projection technique \cite{red}-\cite{must}.

With the mathematical development of the consistent inclusion
of internal symmetries in a statistical thermodynamical
description of quantum gases \cite{red} the idea was applied 
to the colour $ SU(3)_c $ group \cite{gor,elze}.
Therein the group theoretical projection technique was used to project
out colour-singlet representation for a bulk system consisting of 
Quark-Gluon Plasma (QGP) at finite temperature. 
The requirement of imposition of colour-singletness for these systems 
has been found to be of great significance and much work has been done 
using this technique of colour projection \cite{gor2}-\cite{must} . 
Several interesting results were obtained 
but perhaps the most significant was that if one were to compare 
a colour unprojected bulk QGP system with a colour-singlet projected
QGP system then important finite size corrections are introduced 
\cite{gor,elze}.
These finite size corrections arising from the imposition of colour-
singletness disappear as the size and/or temperature of the system 
increases. This was taken to mean that for large-sized QGP systems,
which may have been relevant in the early universe QCD phase transition
scenarios one may automatically assume global colour-singletness 
\cite{mull} of the system without any significant modifications. 
This allowed for the possible existence of large size stable quark 
stars (which were trivially assumed to be colour-singlet \cite{glender}) 
in the early universe QCD phase transition \cite{witt}-\cite{mad}. These 
scenarios continue to dominate the hadronization ideas in the big bang
models \cite{lee}. These ideas have also
been extremely significant in the heavy ion collision scenarios as well
\cite{shaw}.

\newpage

{\bf 2. COLOUR - PROJECTION TECHNIQUE} \\
\vskip 0.3 cm

The orthogonality relation for the associated characters 
$ \chi_{(p,\ q)} $ of the (p, q) multiplet of the group $ SU(3)_c $
with the measure function $ \zeta(\phi, \ \psi) $ is

\begin{equation}
\int_{SU(3)_{c}}^{} d \phi \ d \psi \ \zeta \left( \phi, \psi \right) 
\chi^{\star}_{(p,\ q)}\left( \phi, \psi \right) 
\chi_{(p', \ q')} \left( \phi, \psi \right) \ = \ \delta_{p \ p'} 
\ \delta_{q \ q'}
\end{equation}

Let us now introduce the generating function $ Z^{G} $ as

\begin{equation}
Z^{G}(T, V, \phi, \psi) \ = \ \sum_{p,q} \frac{ Z_{(p, \ q)} }{ d(p, q) }
\chi_{(p, \ q)} ( \phi,\psi )  
\end{equation}
with

\begin{equation}
Z_{(p, \ q)} \ = \ tr_{(p,q)} \left[ 
\exp{ \left( -\beta \hat{H}_{0} \right) } \right]
\end{equation}

\noindent $ Z_{(p,q)} $ is the canonical partition function.
The many-particle states which belong to a given multiplet (p, q)
are used in the statistical trace with the free hamiltonian 
$\hat{H}_0$, d(p, q) is
its dimensionality and $\beta$ is the inverse of the temperature T.
The projected partition function $ Z_{(p, q)} $
can be obtained by using the orthogonality relation for the 
characters. Hence the projected partition function for any representation
(p, q) is

\begin{equation}
Z_{(p, \ q)} \ =  \ d(p, q)
\int_{SU(3)_{c}} d\phi \ d\psi \ \zeta \left( \phi, \psi \right) 
\chi^{\star}_{(p, \ q)}(\phi, \psi) \
Z^G \left( T, V, \phi, \psi \right)
\end{equation}


The characters of the different representations are as follows: \\
\begin{eqnarray}
 \chi_{(1, \ 0)} & = & exp{(2 i \psi/3)} + 
                      2 \ exp{(-i \psi/3)} \ cos(\phi/2) \\
  \chi_{(0, \ 1)} & = & \chi_{(1, \ 0)}^{\star} \\
  \chi_{(1, \ 1)} & = & 2 \ + \ 2 \ \left[ cos\phi 
    +  cos\left( \phi/2 + \psi \right) 
    +  cos\left(-\phi/2 + \psi \right) \right] \\
  \chi_{(2, \ 2)} & = & 2 \ + \ 2 \ \left[ cos \phi  
    +  cos (3\phi/2) cos (\phi/2) \right] + \nonumber \\
               &   & 2  \ \left( 1 + 2 \ cos\phi \right) 
       \{ cos\left( \phi/2 + \psi \right) 
    +         cos\left(-\phi/2 + \psi \right) 
            + cos 2 \psi + \left( 1/2 \right) cos \phi 
       \} 
\end{eqnarray}
The expressions of the generating function used in (4) is 

\begin{equation}
 Z^G(T, V, \phi, \psi) 
 \ = \ tr\left[exp(-\beta \hat{H}_{0}
         + i \phi \hat{I}_z + i \psi \hat{Y}) \right] 
\end{equation}

where $\hat{I}_{z}$ and $\hat{Y}$ are the diagonal generators of the 
maximal abelian subgroup of $SU(3)_c$. 
 Our plasma consists of light spin 1/2 (anti) quarks in the 
(anti) triplet representation (0, 1) and (1, 0) respectively, and massless
spin one gluons in the octet representation (1,1). Note that the 
non-interacting hamiltonian $\hat{H}_0$ is diagonal in the occupation-number
representation. In the same representation one can write the charge 
operators $\hat{I}_z$ and $\hat{Y} $ as linear combinations of particle-number 
operators. Hence $Z^G$ can be easily calculated in the occupation-number 
representation. With an imaginary `chemical potential' this is just like a 
grand canonical partition function for free fermions and bosons. One obtains

\begin{eqnarray}
Z^G_{quark} \ = \ \prod_{q = l, m, n} \prod_{k} 
   \left[ 1 + exp{( -\beta\epsilon_k - i\alpha_q) } \right] 
   \left[ 1 + exp{( -\beta\epsilon_k + i\alpha_q) } \right] \\
Z^G_{glue} \ = \ \prod_{g = \mu, \nu, \rho, \sigma} \prod_{k} 
   \left[ 1 - exp{(-\beta\epsilon_k+i\alpha_g)}  \right]^{-1}
   \left[ 1 - exp{(-\beta\epsilon_k-i\alpha_g)}  \right]^{-1}
\end{eqnarray}

\noindent Here the single-particle energies are given as 
$\epsilon_k$. For (1, 0), (0, 1) and (1, 1) multiplets, 
the eigenvalues of $ \hat{I}_z $ and $ \hat{Y} $ gives the expression for 
different angles as:

\begin{eqnarray}
\alpha_l \ = \ (1/2) \phi + (1/3) \psi, \ \alpha_m \ = \ 
(-1/2) \phi + (1/3) \psi, \ \
 \alpha_n \ = \ (-2/3) \psi \\
\alpha_{\mu} \ = \alpha_l-\alpha_m, \ \alpha_{\nu} = \alpha_m-\alpha_n, \
  \alpha_{\rho} = \alpha_l-\alpha_n, \ \alpha_{\sigma} = 0
\end{eqnarray}

\noindent We neglect the masses of the light quarks. 
At large volume the spectrum of single particle becomes a quasi-continuous 
one and $\Sigma ... \rightarrow V/({2\pi})^{3} \int d^{3}p ... $ Then one gets

\begin{equation}
Z^G (T, V, \phi, \psi) \ = \
Z_{quark}^G (T, V, \phi, \psi) \ Z_{glue}^G (T, V, \phi, \psi) 
\end{equation}


This then enables us to obtain the partition function for any 
representation ie. $Z_{(p, \ q)}$. 
One may thus obtain any thermodynamical quantity of interest for a particular 
representation. For example the energy 
\begin{equation}
 E_{(p, \ q)} \ = \ T^2 \frac{\partial}{\partial T} \ln Z_{ (p, \ q) } .
\end{equation}

\vskip 1.5 cm

{\bf 3. RESULTS } \\
\vskip 0.3 cm

Work was done earlier by several groups to impose colour-singletness on
the system \cite{elze, gor2}. Note that in these calculations perturbative
interactions had been neglected. But this may not be a bad apprximation
especially at high temperatures. The most dramatic consequence of the colour
interaction is to cause global colour-confinement of quarks and gluons
and this is automatically taken care of by restricting the partition 
function to colour-singlet states \cite{mull}. This perhapes may be taking 
care of a major chunk of non-perturbative aspect of QCD interaction.
It was found that 

\begin{equation}
E_{(0, \ 0)} \ = \ E_{ 0 } \ + \ E_{corr} 
\end{equation}

\noindent where $ E_{0} $ was the unprojected energy (ie. with no colour 
restriction whatsoever) given by

\begin{equation}
 E_0 \ = \ 3 \ a_q \ V \ T^4 
\end{equation}

\noindent with  $ a_q = (37 \ \pi^2/90) $
and $E_{corr}$ was the correction introduced due to the imposition of 
colour-singletness. They found that \cite{elze, gor2}
$E_{corr}$ was significant only for the finite size i.e. when $TV^{1/3}$
was small ( $<2$) and vanished when $TV^{1/3}$ became large ($>2$).
This would mean that colour-singletness restriction only affects for size 
say $\sim 1.0 $ fm for T = 160 MeV while for large size and higher 
temperatures one need not perform explicit colour projection calculation
because the consequent corrections are negligible therein \cite{elze, gor2}.
But below we shall show that this is not the whole story.

In this paper we project out different representations like
octet (1, 1), 27-plet (2, 2) etc. on these QGP. The idea is 
that for ground state one knows that the singlet state 
is bound and the higher representations are expelled to
infinite energies \cite{mars}. Also for the ground states the role
of the higher representations is also quite well studied
\cite{afsar, bra}. The point to be emphasized is that the role of global 
colour-singletness at high temperatures is only an assumption and has never 
been explicitely demonstrated even in a model calculations. 
Here we would like to study the basis of this assumption
and also the role, if any, of higher representations like octet, 27-plet etc. 

Let us look at octet, 27-plet etc. projection. For simplicity we take
$\mu = 0$ case with 2 flavours. We plot in Fig. 1

\begin{equation}
D_{(p, \ q)}^{eff} \ = \ E_{(p, \ q)}/E_0 \ = \ 1 \ + \ E_{(p, \ q)}^{corr}/E_0
\end{equation}

Also shown is $D_{(0, \ 0)}^{eff}$ as obtained earlier by other
groups \cite{elze}. 

\vskip 1.5 cm

{\bf 4. DISCUSSION} \\
\vskip 0.3 cm

It is interesting to note from Fig. 1 that for large values 
of $TV^{1/3}$ all representations ; singlet, octet, 27-plet etc are degenerate 
with the unprojected energy. For sufficiently large QGP and sufficiently high 
temperatures all the energies of all the representations are found to be
degenerate. There is nothing which favours the colour-singlet 
representation over the colour-octet at high temperatures. 
Note that this result could directly be seen
from the expression $Z_{(p, \ q)}$ (eq.(3)) which for the continuum 
approximation for sufficiently large volume and temperature becomes 
independent of representation. Hence energies for each colour representation 
are degenerate with each other for large $ TV^{1/3} $.
As the neglect of perturbative interactions is justified at high enough 
temperatures, this important result seems to be quite model independent.

Note that the quarks and gluons are non-interacting in our model. While
this is justified at high temperatures, the neglect of interactions may
not be justified at low temperatures. However as one of the most dramatic
effect of colour interaction is ensuring colour-singletness on the bulk
system it is already taken care of in our model \cite{mull}.
So perhapes a significant portion of the non-perturbative effect may 
already be there in our model calculations. From Fig. 1 note that for small 
$TV^{1/3}$ values the octet and the 27-plet energies shoots up. 
Though gauge interactions are believed to be
essential to show confinement in QCD, what we note here is that our
projection technique at even low temperatures is able to discriminate between
the singlet and the octet states etc.

The $\mu \neq 0$ case may be significant for giving finer details
\cite{lina}, however, we do not expect any significant change in our
basic results \cite{elze}.
Fig. 2 gives $\mu = 0$ result for zero flavour, 2-flavour
and 3-flavour for (0, 0), (1, 1) representations.

We have found the degeneracy of the states for large $ TV^{1/3} $
for each colour representation for $\mu = 0$ to be true for
0, 2, 3 flavours etc. So it is independent of number of flavours.
We find it occurs for the $ \mu \neq 0 $ case also \cite{lina}. 
The fact that the colour singlet representation gets
favoured over the octet representation etc. at low temperature, group 
theoretically it indicates that for composite systems
$SU(3)_c$ is a good symmetry only for small temperatures and above the 
QCD phase transition temperature (actually $ TV^{1/3} $) gets submerged 
into a larger group. We would like to emphasize that this is definitely true
at high enough temperatures. This larger group is 
$ U(18)_q \otimes U(18)_{\bar{q}} $ for 3-flavours,
$ U(12)_q \otimes U(12)_{\bar{q}} $ for 2-flavours,
$ U(6)_q \otimes U(6)_{\bar{q}} $ for 0-flavours. 
The subgroup structure for say the 2-flavour case is
\begin{equation}
 U(12)_q \otimes U(12)_{\bar{q}} 
              \supset SU(12)_{csf}
              \supset SU(6)_{cs} \otimes SU(2)_f
              \supset SU(3)_c \otimes SU(2)_s \otimes SU(2)_f
\end{equation}
where the subscripts 'csf' stand for colour, spin and flavour respectively.
It is only below the QCD phase transition temperature that $SU(3)_c $
becomes a good symmetry for the composite system. For the composite system,
at high temperatures $ SU(3)_c $ does not remain the relevant symmetry  
and gets submerged into 
$ U(12)_q \otimes U(12)_{\bar{q}} $ (for 2-flavours).
As transition from hadronic matter to quark-gluon matter is a transition
from ``local" colour confinemnt to ``global" colour confinement \cite{mull},
in view of our result it also implies going from the group $SU(3)_c$ to
$ U(12)_q \otimes U(12)_{\bar q}$.

We would like to remind the reader that in the quark model in the static 
limit it was found fruitful to look at $SU(3)_f$ as submerged in
$SU(6)_{fs}$ as $ SU(6)_{fs} \supset SU(3)_f \otimes SU(2)_s $. In the 
same spirit we have found here that at high temperatures and/or large
size the $SU(3)_c$ gets deflated and submerged in a larger group e.g. 
$U(12)_q \otimes U(12)_{\bar q}$. We would like to emphasize that at 
sufficiently high temperatures and/or large size this conclusion is
model independent.

Note that the currently favoured scenario for the early universe above
the QCD and the electroweak phase transition temperatures, the unbroken
group is believed to be $ SU(3)_c \otimes SU(2)_L \otimes U(1)_Y$.
Our work in this paper casts doubt in this view. Globally it is not simply 
$SU(3)_c$ but $ U(12)_q \otimes U(12)_{\bar q}$ in the early universe.
Ideas of extensions beyond the standard model have to be reviewed in the
light of this result.

\newpage

{ \large {\bf Figure Captions } }
\vskip 2 cm
{\bf Fig. 1 : }
$ D_{eff} $ (see text) for the colour representations singlet, 
octet and 27-plet (with two flavours) as a function of $ TV^{1/3} / \hbar c $
\vskip 3 cm
{\bf Fig. 2 : }
$ D_{eff} $ for the colour representations singlet and octet 
as a function $ TV^{1/3} / \hbar c $  
for different number of flavours ; 0, 2 and 3.

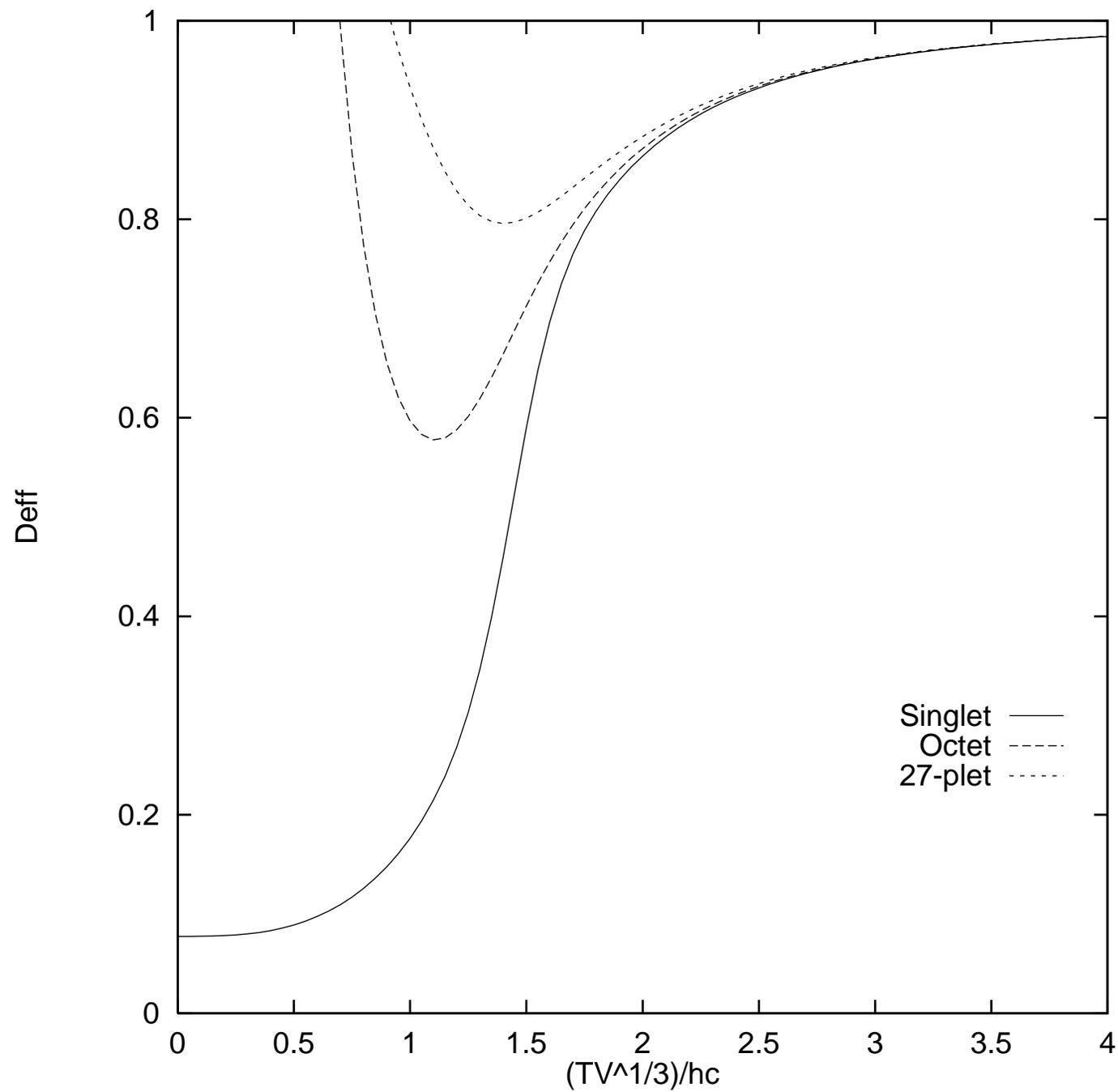
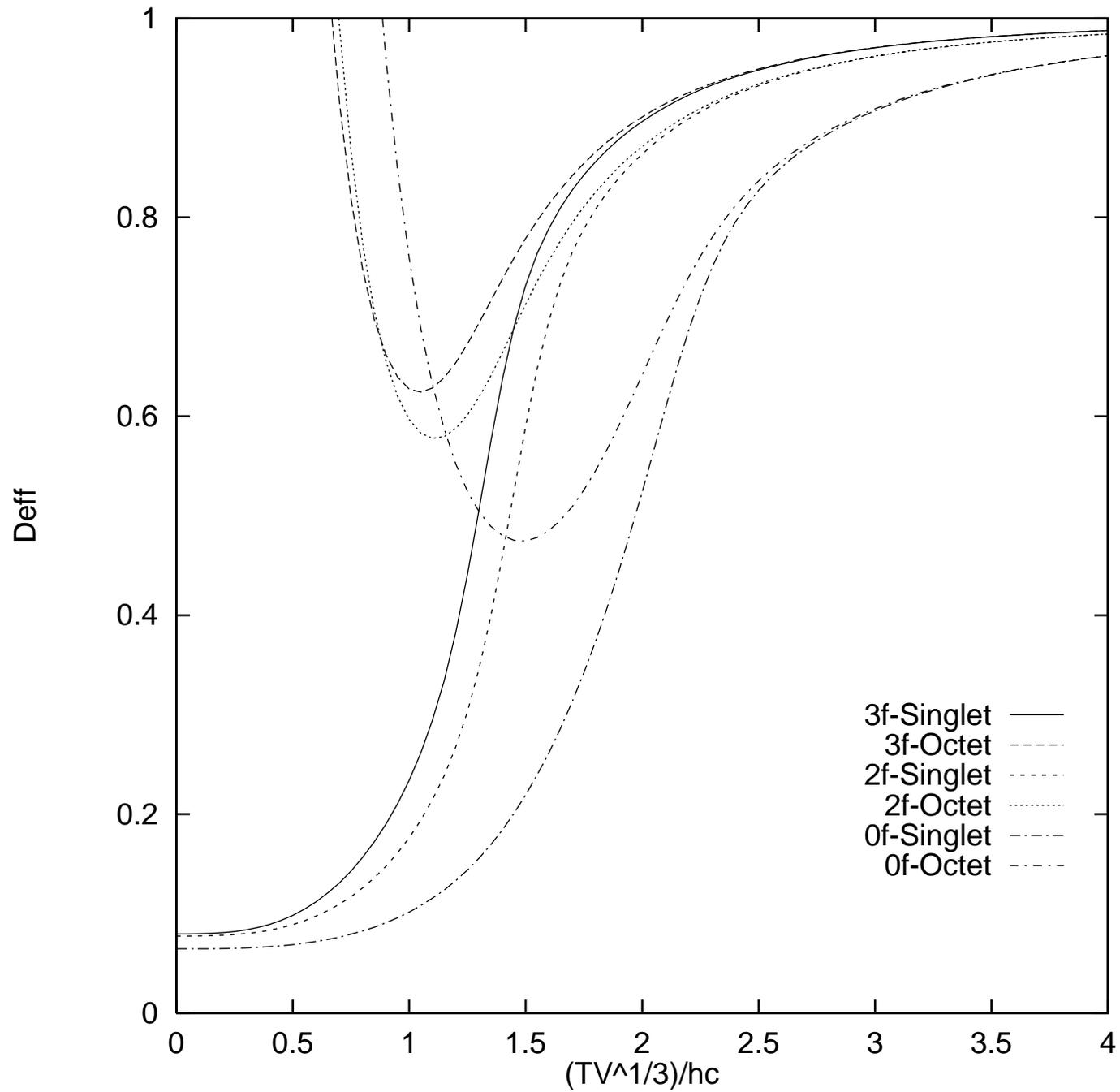
\end{document}